\documentclass[prl,preprint,showpacs,preprintnumbers,amsmath,amssymb,floatfix]{revtex4}
\usepackage{graphicx}
\usepackage{natbib}
\usepackage{bm}
\usepackage{mathrsfs}
\usepackage{verbatim}
\usepackage{wasysym}

\def\hf{{\frac{1}{2}}}

\begin{document}

\title{Lorentz Covariant Lattice Gauge Theory}
\author{Timothy D. Andersen}
%\affiliation{Dept. of Mathematics and Statistics, Old Dominion University, 5110 Hampton Blvd., Norfolk, VA, USA}
%\ead{andert@alum.rpi.edu}
\email{andert@alum.rpi.edu}
\date{Received: date / Accepted: date}
\begin{abstract} Lattice gauge theory's discretization of spacetime suffers from a drawback in that Lorentz covariance is lost because the axes of the lattice create preferred directions in spacetime. Smaller and smaller lattice spacings decrease the effect but fail to eliminate it completely. It has been argued recently that detecting such a set of preferred directions or similar constraints would indicate whether the universe itself has an underlying lattice, i.e. the digital universe hypothesis. In this paper, I demonstrate a technique for accomplishing lattice gauge theory simulations while maintaining exact Lorentz covariance by replacing the lattice with a lattice graph such that the metric is defined as a discrete, Lorentz covariant matrix potential over the graph rather than a metric over an underlying manifold. This technique eliminates the symmetry violation of standard lattice gauge theory and suggests that, even in a digital universe, Lorentz covariance can still hold.
\end{abstract}

%\submitto{CQG}
\maketitle
%\flushbottom

Standard lattice gauge theory was first proposed by Wilson in his seminal paper \cite{Wilson:1974} on the confinement of quarks. In his computational paradigm, the spacetime manifold is discretized by a hypercubic lattice. Actions for theories such as QCD are computed by summing over the plaquettes  (closed loops formed by four edges) of the lattice via the discrete Yang-Mills action for the plaquette. The form of the action as a product of symmetry group members maintains the symmetry group of the theory, e.g., SU(3). The symmetry group of spacetime itself, however, is not maintained. In other words, rotations, boosts, and translations of the lattice cause the potentials and sources and hence the action on the lattice to change. If the universe has Poincar\'e symmetry (or de Sitter symmetry), then lattice gauge theory only approximates this symmetry for sufficiently small lattice spacings. Recently, it has been suggested that the universe may have an underlying lattice, the so-called ``computer simulation'' universe \cite{Bostrom:2003} where real phenomena are the result of sophisticated lattice gauge computations, and that this could be detected by violations of the Greisen--Zatsepin--Kuzmin (GZK) limit \cite{Beane:2012}, resulting in a breakdown of special relativity at high cosmic ray energies. The authors, however, admit that computational paradigms ``beyond our comprehension'' could be used to eliminate the violation they are looking for. Observations that constrain violations of Lorentz invariance include astrophysical \cite{Jacobson:2003} and neutron electic dipole measurements\cite{Altarev:2011} but these are at much lower energies than the GZK. In this paper, I show that lattice gauge simulations need not violate Lorentz symmetry and that the computational method to do so is well within the realm of human comprehension. Therefore, even in a computer simulation universe, Lorentz symmetry is not necessarily violated.

Lattice gauge theory violates Lorentz invariance because the lattice is embedded in a mathematical abstraction, the manifold, which represents ``real'' spacetime \cite{Zee:2003}. The manifold is continuous and differentiable and potential and source fields used in physics computations are defined as differentiable functions on it. Lattice gauge computations represent the lattice  explicitly as sequences of vertices connected by edges while the existence of the manifold is implicitly understood. Because the computational physicist chooses the lattice's orientation within the manifold, a non-physical ``preferred'' coordinate system is imposed on spacetime. For computational purposes, as long as the lattice spacings are small, the coordinate systems should not affect the computations. This is not guaranteed in all cases, however, because, as in the growth of crystals, the cubic structure can affect the ``macroscopic'' scale under the right circumstances. As lattice gauge computations become more complex with greater computing power, e.g. Petaflops (such as with DOE's Jaguar supercomputer and the upcoming Titan) and Exaflops likely to be achieved in the next decade, problems may arise because of the underlying cubic crystalline structure of the lattice. (DeGrand, et. al \cite{Degrand:2006} is a good reference for lattice gauge methods for QCD.)

To solve this problem, an alternative to the standard lattice approach is to define the lattice with no underlying manifold so that it is just a graph. To be clear, when there is no underlying manifold, the edges and vertices have no position (or length) whatsoever, and rotations, boosts, and translations of the lattice itself have no meaning. Potential and source functions become discrete functions on the vertices, which relate to one another via the edges of the graph. Hence, the cubic crystalline structure of the former preferred coordinate system no longer exists because there is no coordinate system. 

Removing the manifold creates several new problems. There is now no coordinate system and hence no way to measure distances between event points that occur at the vertices, which is essential to quantum field theory. There is also no differentiability, which means that the continuum limit Lagrangians that contain derivatives cannot be obtained from the lattice as it now stands. Intuitively, it seems like a manifold is essential to defining a theory that describes quantum behavior, but, as I show in this paper, it is not. Both of these problems are solved, instead, by defining a gauge theory on the lattice graph such that Poincar\'e or de Sitter covariance is conserved, and the continuum limit recovers the standard Yang-Mills action. The gauge theory ensures that the lattice reflects all the symmetries of a flat manifold without the manifold. No curved spacetime manifolds are considered.

First, an one dimensional example distinguishes between the standard approach and the approach I take below. Consider the real number line on which is defined a metric. For an arbitrary continuous, differentiable function defined on the real numbers, $f: \mathbb{R}\rightarrow\mathbb{R}$, a translation invariant functional $S$ on $f$, $S[f] = \int dx L[f(x)]$ can by approximated by evaluating it over a lattice $\mathcal{L}_0 =\{\dots,-2\epsilon,-\epsilon,0,\epsilon,2\epsilon,\dots\}$, $S[f]\approx \sum_{k=-\infty}^\infty \epsilon L[f(k\epsilon)] = S_\epsilon[f]$. If we translate the lattice: $\mathcal{L}_0=\{\dots,-2\epsilon,-\epsilon,0,\epsilon,2\epsilon,\dots\} \rightarrow\mathcal{L_\delta}=\{\dots,-2\epsilon+\delta,-\epsilon+\delta,\delta,\epsilon+\delta,2\epsilon+\delta,\dots\}$ by an amount $\delta$, the approximation to the functional is, for arbitrary $f$ and $S$, slightly different than the untranslated approximation: $S_{\epsilon,\delta}[f] = S_{\epsilon}[f] + \sigma$ for some small difference $\sigma$. Therefore, the functional approximated on the lattice is not translation invariant, unlike the functional on the complete real line. In the continuum limit as the lattice spacing vanishes, however, translational invariance is recovered: $S_{\epsilon\rightarrow 0,\delta}[f] = S_{\epsilon\rightarrow 0}[f]=S[f]$. 

In higher dimensions, we also have boosts and rotations in addition to translations, but the principle is the same: discretizing over a lattice within a spacetime manifold eliminates the natural covariance of the manifold.

The translation invariance can be added by redefining the lattice as a graph. The lack of translation invariance started because we assumed the lattice was embedded in the real number line. Suppose we eliminate the real number line and define, instead, an infinite graph where each vertex has exactly two neighbors. Label the vertices $V=\{\dots,y_{-2},y_{-1},y_{0},y_1, y_2,\dots\}$. Define a function $g_k:V\rightarrow\mathbb{R}$ on each vertex. This function $g_k$ is a distance function and gives the distance between $y_{k-1}$ and $y_k$. Let $f_k:V\rightarrow \mathbb{R}$ be an arbitrary function defined on the lattice. The approximation to the functional above is given by $S_{g}[f] = \sum_{k=-\infty}^\infty g_k L[f_k]$ and is translation invariant because $g_k$ is translation invariant. Let $g_k=\epsilon$ and $f_k = f(k\epsilon)$ and $S_g[f] = S_{\epsilon}[f]$. If $f_k$ satisfies \[\lim_{g_k\rightarrow 0} \frac{f_k - f_{k-1}}{g_k} \] finite such that it is differentiable in the limit, then $S_g[f]\rightarrow S[f]$ as $\epsilon\rightarrow 0$. Although this example is too simple in one dimension to demonstrate the gauge covariance of $g_k$ which will become evident in higher dimensions, it illustrates the main feature of the approach, which is to eliminate the metric on the underlying manifold in favor of a distance function that is tied to the lattice.

Extending this approach to higher dimensions is much trickier than simply defining a metric function $g_{jk}$ that gives the distance between every pair of vertices $y_j$ and $y_k$ because it does not provide a way to represent coordinate transformations. In one dimension, there is no coordinate freedom in the metric. The distance function $g_k$, which, in our special case, was simply $\epsilon$, is a scalar field and necessarily invariant. In higher dimensions, one $g_{jk}$ and another $g'_{jk}$ may be equivalent under a coordinate transformation, e.g. rotation, because they covary. In the standard, lattice-in-a-manifold approach, the manifold takes care of that covariance via transformation rules for differentiable manifolds. On a lattice with no manifold, different rules apply which are, nevertheless, equivalent in the continuum limit. These are the rules of gauge covariance:

(A note on subscripts: The beginning letters of the latin alphabet $a,b,c,\dots,h$ represent 4-vectors. These letters range over $0,1,2,3$. I will use $y_4$ to refer to the additional, 5th dimension that completes de Sitter group vectors.)

First, we must define the gauge theory as in the 1-D example but now for a 3+1-D universe. Let time be imaginary $t\rightarrow i\tau$ so that spacetime is Euclidean and boosts become rotations, and define the coordinate system as de Sitter. (This is for simplicity. The following theory is achieved in the same way for the Poincar\'e symmetry group.) On a four dimensional, infinite lattice graph such that every vertex has eight neighbors, label each vertex with five numbers $y_0,y_1,y_2,y_3,y_4$. These labels become the coordinate system once the metric is defined. The fifth component, $y_4$, is constant under the Poincar\'e group, with $y_4=1$ for all vertices, but can change under the de Sitter group. For three vertices, ${\bf y}$ neighbor to ${\bf z}$ neighbor to ${\bf w}$, let $U_{{\bf yw}}({\bf z})$ be a {\em small} rotation matrix with $\epsilon>0$ the small parameter. The matrix $U_{{\bf yw}}=\exp(\epsilon A_a)$ can be split into two real valued parts: $G_{ab}$ and $H_{abc}$ as a sum of generators:
\begin{equation}
A_{a} = G_{ab}V_b + \hf H_{abc}M_{bc},
\label{eqn:matrixpot}
\end{equation} where $V_b$ are generators for rotations in the $y_b$-$y_4$ plane and $M_{bc}=-M_{cb}$ are generators for SO(4) for rotations in the $y_b$-$y_c$ plane where $a,b,c=0,1,2,3$. Each of the ten generator matrices, $V_b$ and $M_{bc}$, is a $5$ by $5$ matrix. 

A note on terminology: These matrices generate rotations in all 10 planes of a five dimensional space, but the $V_b$ generators can be considered ``translations'' for sufficiently small rotations, where the de Sitter group approximates the Poincar\'e group. Therefore, in the following, I refer to the transformations under $V_b$ generators as ``translations'' with the understanding that, in the case of the de Sitter group, these are actually rotations with respect to a fifth coordinate axis. I refer to $G_{ab}$ as the {\em metric potential} and $H_{abc}$ as the {\em torsion potential} because of their roles in determining distances and torsions (twists in parallel transport of vectors). In general, torsion will be zero.

Now, we need to extract a metric from our ``spin-1'' matrix potential, $A_a$. To linear order in $\epsilon$ the rotational matrix splits into component form,
\begin{equation}
U_{yw} = I + \epsilon A_a = I + \epsilon (G_{ab}V_b + \hf H_{abc}M_{bc}).\nonumber
\end{equation} 

Using this transformation, we can define the coordinate system in terms of the transformation potentials. Let ${\bf y} = (y_0,y_1,y_2,y_3,y_4)$ and ${\bf w} = (w_0,w_1,w_2,w_3,w_4)$, then
\begin{equation}
{\bf w} = [I + \epsilon (G_{ab}V_b + \hf H_{abc}M_{bc})]{\bf y}.
\label{eqn:fivebyfive}
\end{equation}  The vector ${\bf y}$ experiences, on its way to becoming ${\bf w}$, a transformation (translation or rotation) by passing through ${\bf z}$. The formulas $\mathrm{tr}\, V_{a}V_b = \delta_{ab}$ and $\mathrm{tr}\, M_{ab}M_{cd} = \delta_{ac}\delta_{bc}$ \cite{Zee:2003} eliminate the generator matrices in \ref{eqn:fivebyfive}:
\begin{equation}
w_a \equiv \hf \epsilon H_{abc}y_c + \epsilon G_{ab} + y_b.
\label{eqn:coord}
\end{equation} The first term in the right hand side of \ref{eqn:coord} represents the torsion or twisting of a vector as it travels from ${\bf y}$ through ${\bf z}$ to end at ${\bf w}$ and is included for the sake of completeness but will be set to zero later. The second term on the right hand side represents the amount of translation. (Note: if $G_{ab}({\bf z})=0$ then there is no translation and the two lattice points are in the same location.) 

To see how the above potentials describe a coordinate system without reference to a manifold, consider an arbitrary matrix potential component $B_1(y_a)$ as an example. Label the eight edges attached to vertex $y_a$ in pairs as $1,-1, 2,-2, 3,-3,$ and $4,-4$. Then if $B_1(y_a)\geq 0$, its direction corresponds to edge $1$. If $B_1(y_a)<0$, we have edge $-1$. The edge's direction in coordinate space is defined by \ref{eqn:coord}. Likewise, the position $y_a$ relative to its neighbors is defined by \ref{eqn:coord}.

Now, we need to obtain a standard Euclidean coordinate system. Given that  spacetime is flat (gravity is insignificant), there exists a gauge transformation $G_{ab}\rightarrow G'_{ab}$ such that the metric potential is Euclidean: $G'_{ab}=\delta_{ab}$ and $H'_{abc}=0$, where, for the $SO(5)$ transformation $O(x)$ such that $O^{-1}=O^T$, the gauge transformation of the matrix potential in \ref{eqn:matrixpot} is 
\begin{equation}
A'_{a} = OA_aO^{T} - (\partial_a O)O^{T}.
\end{equation} This equidistant Cartesian coordinate system corresponds to a lattice where each vertex is a distance $\epsilon/2$ from each of its eight neighbors. Note: the distance between two non-neighboring event points is undefined and is not necessary to lattice gauge simulations. Instead, relationships between non-neighboring points are given indirectly by probabilistic correlations, which the lattice gauge simulation samples. In the classical limit $\hbar\rightarrow 0$, these correlations lead to the ``macroscopic'' notion of distance, but this is beyond the scope of lattice gauge theory.

\begin{figure}
\centering
\includegraphics[width = 0.35\textwidth]{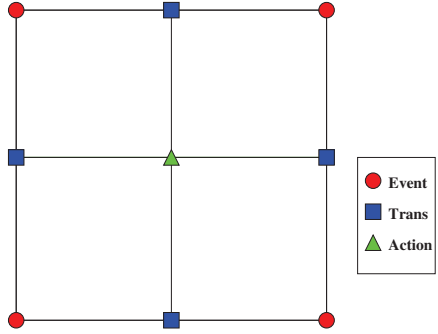}
\caption{The plaquette consists of three kinds of points: event points, transition points, and an action point. This ensures that every quantity relevant to the lattice gauge computation is defined on a vertex and avoids having the action, in particular, appearing to be defined beyond the graph.}
\label{fig:plaquette} 
\end{figure}

Now, we define our lattice gauge theory. The metric and torsion potentials are defined on the vertices between the event points. There are three kinds of vertices: event points where sources are defined, transition or transformation points where gauge potentials are defined, and action points where actions are defined (Fig. \ref{fig:plaquette}). This definition is equivalent to that given in \cite{Zee:2003}.

In the Wilsonian derivation of any Yang-Mills theory, e.g. QCD, the metric appears from summing over plaquettes \cite{Wilson:1974}\cite{Zee:2003}. Given an action as a sum of products of transformation matrices,
\begin{equation}
S_\epsilon = \mathrm{tr}\,\sum_{\textrm{plaquettes}} U_iU_jU_kU_l,
\label{eqn:ymaction}
\end{equation} where each plaquette has sides $i,j,k,l$ in a closed loop and $U_i,U_j,U_k,U_l\in SU(N)\otimes SO(5)$. Each matrix has an SU(N) submatrix in the upper left corner and a constant SO(5) submatrix in the lower right corner. Eqn. \ref{eqn:ymaction} is invariant under a change of coordinate gauge $y_a\rightarrow \chi^b{}_a y_b + \omega_a$ because, by \ref{eqn:coord}, the potential values defined on the vertices are fixed irrespective of how the coordinate system changes.

Since the method is restricted to straight-line coordinates such as Cartesian coordinates and global, i.e. constant, SO(5) transformations, the usual gauge transformation reduces to an ordinary rotation transformation:
\begin{equation}
A'_{a} = OA_aO^{T},
\end{equation} such that, in real time, ordinary Lorentz transformations transform the metric:
\[
G'_{ab} = \Lambda^c_a\Lambda^d_bG_{cd},
\] and torsion remains zero: $H_{abc} = 0$. Thus, in equidistant Cartesian coordinates, $G_{ab} = \eta_{ab}$. Because the SO(5) part of each $U$ matrix in \ref{eqn:ymaction} is constant, it also drops out of the final action leading to a continuum limit identical to standard Yang-Mills theory. By \ref{eqn:coord}, the metric potential $G_{ab}$ appears everywhere the manifold metric usually appears.

\begin{figure}[h]
\centering
\includegraphics[width = 0.45\textwidth]{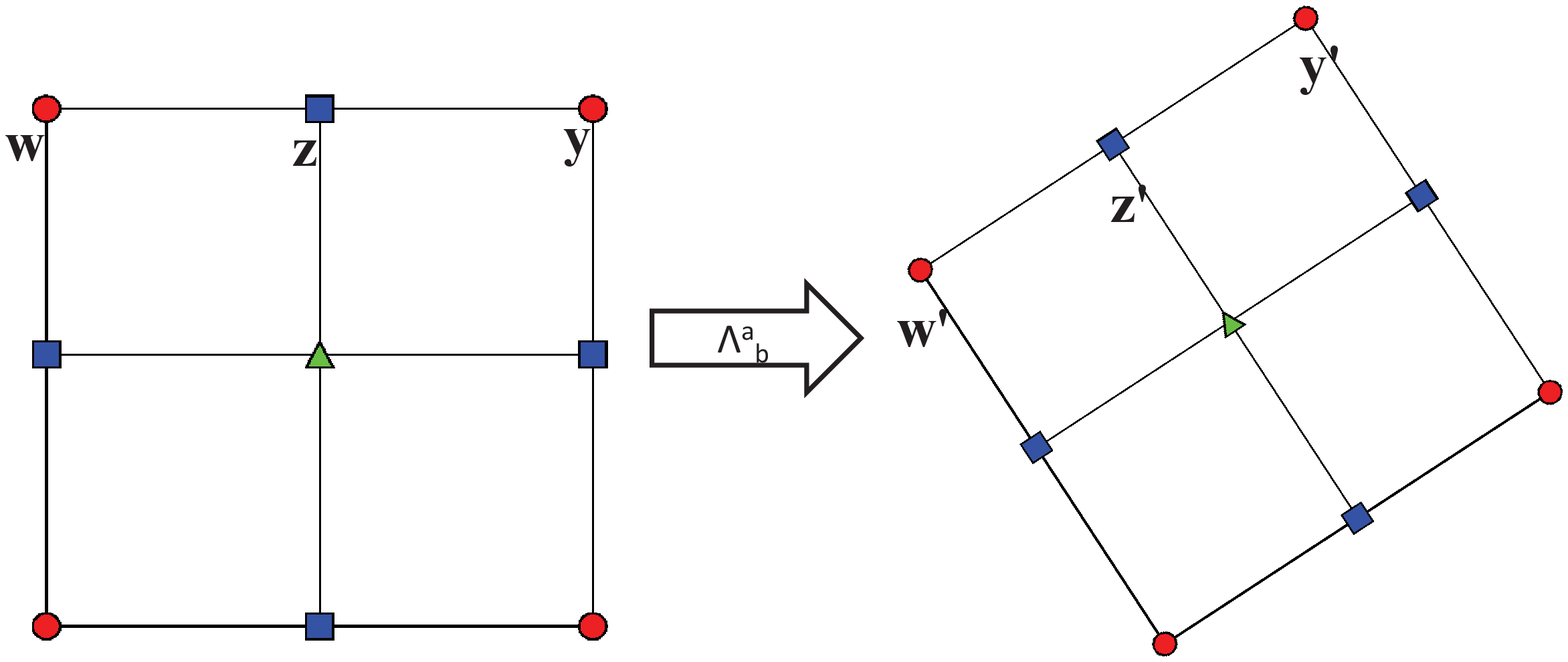}
\caption{The plaquette with a background manifold transitioning from frame $K$ to $K'$.}
\label{fig:plaquette1} 
\end{figure}

\begin{figure}[h]
\centering
\includegraphics[width = 0.45\textwidth]{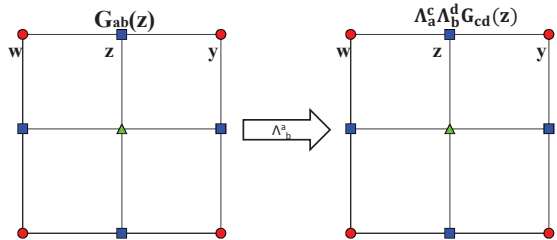}
\caption{The plaquette without a background manifold transitioning from frame $K$ to $K'$.}
\label{fig:plaquette2} 
\end{figure}

Because $G_{ab}$ is the same as the standard metric for a flat spacetime, lattice gauge simulations with the modified method will be identical to those with the standard method for a given orientation of the lattice. If the lattices are transformed by a Lorentz transformation, however, a difference between the two will arise. The difference between the background manifold and no background manifold methods is illustrated in Figures \ref{fig:plaquette1} and \ref{fig:plaquette2} for a transition between two frames which are rotated with respect to one another by a transformation $\Lambda^a_b$ from a frame $K$ to $K'$.


This paper has shown a subtle difference between a lattice embedded in a manifold and a lattice with no manifold. In the absence of gravity, there is no difference between the two in the continuum limit (assuming differentiability holds on all functions over the lattice), but, before the continuum limit is taken, only one is Lorentz covariant. This shows that even in a so-called digital universe of discrete lattice points, observables can be absolutely conserved under continuous Lorentz transformations. Further work would attempt to show adverse effects from Lorentz symmetry violations in lattice gauge simulations and also extend to curved spacetimes.

\bibliography{sg2}

\end{document}